# Symplectic Calculation of Lyapunov Exponents

**Salman Habib**[†] **and Robert D. Ryne**[⋆]

[†]*T-6, Theoretical Astrophysics*
*and*
*T-8, Elementary Particles and Field Theory*
*Theoretical Division*
*Los Alamos National Laboratory*
*Los Alamos, NM 87545*

[⋆]*AOT-1, Accelerator Physics and Special Projects*
*Accelerator Operations and Technology Division*
*Los Alamos National Laboratory*
*Los Alamos, NM 87545*

**Abstract**

The Lyapunov exponents of a chaotic system quantify the exponential divergence of initially nearby trajectories. For Hamiltonian systems the exponents are related to the eigenvalues of a symplectic matrix. We make use of this fact to develop a new method for the calculation of Lyapunov exponents of such systems. Our approach avoids the renormalization and reorthogonalization of usual techniques. It is also easily extendible to damped systems. We apply our method to two examples of physical interest: a model system that describes the beam halo in charged particle beams and the driven van der Pol oscillator.

e-mail:
habib@eagle.lanl.gov
ryne@lanl.gov

Chaotic dynamics has been observed in a wide variety of systems, including biological systems, weather models, mechanical devices, plasmas, and fluids, to name a few. In the chaotic regime these systems exhibit exponential divergence of initially nearby trajectories, as embodied by the Lyapunov exponents of the system [1]. Over the past two decades or so there has been "intense activity" [2] directed toward the computation of these exponents resulting in several different numerical approaches [1][3][4]. The two difficulties associated with the computation of Lyapunov exponents are: (1) exponential growth of the separation vector (between the fiducial and nearby trajectories) and (2) the exponential collapse of initially orthogonal separation vectors onto the direction of maximal growth. Most conventional methods overcome these hurdles by intermittent numerical rescaling and reorthogonalization (through, *e.g.*, the Gramm-Schmidt procedure [3]). Many chaotic systems are Hamiltonian or they can be transformed into a Hamiltonian system by suitable manipulations. However, none of the above general methods are designed to take advantage of this fact.

The dynamics of classical Hamiltonian systems has an underlying symplectic structure [5]. In recent years symplectic methods have been applied with great success to classical dynamical problems. The field of accelerator dynamics has been revolutionized by the introduction of nonlinear symplectic maps as represented by Lie transformations [6][7]. Very long time integration of charged particle and planetary systems has been aided by the development of high order symplectic integration algorithms [8]. In this Letter we exhibit a symplectic map-based approach towards the calculation of Lyapunov exponents. As shown below, this approach obviates *analytically* the need for rescaling and reorthogonalization in the numerical computation of the exponents. We have successfully applied our method to several systems, including the Duffing oscillator, the damped driven pendulum, the driven double-well system, the beam halo system, and the van der Pol oscillator (details will be presented elsewhere [9]). In this Letter, we will briefly describe our general approach and expose in some detail the latter two examples.

Consider a 2-$m$ dimensional continuous-time dynamical system governed by the equations

$$\frac{d\mathbf{z}}{dt} = \mathbf{F}(\mathbf{z}, t), \tag{1}$$

where $\mathbf{z} = (z_1, z_2, \cdots, z_{2m})$ and similarly for $\mathbf{F}$. Let $\mathbf{z}_0$ denote some given fiducial trajectory, and suppose we wish to study nearby trajectories. To do this, define deviations from the fiducial trajectory by letting $\mathbf{Z} = \mathbf{z} - \mathbf{z}_0$, and linearize the above equations. The new set of equations for the deviation variables is

$$\frac{d\mathbf{Z}}{dt} = \mathbf{DF}(\mathbf{z}_0, t) \cdot \mathbf{Z}. \tag{2}$$

Our approach can be used whenever this linearized set of equations can be derived from a Hamiltonian. From now on we will suppose that this is the case, and that we can write $\mathbf{Z} = (q_1, q_2, \cdots, q_m, p_1, p_2, \cdots, p_m)$, where $q_i$ and $p_i$ denote canonically conjugate coordinates and momenta, respectively. It follows that

$$\frac{d\mathbf{Z}}{dt} = -\{H, \mathbf{Z}\}, \tag{3}$$

where $\{,\}$ denotes the Poisson bracket, and where $H$ is a homogeneous quadratic polynomial in the $q_i$ and $p_i$. A system such as this is governed by a symplectic matrix $M$ that maps the initial variables $\mathbf{Z}^{in}$ into time-evolved variables $\mathbf{Z}(t)$,

$$\mathbf{Z}(t) = M(t)\mathbf{Z}^{in}. \tag{4}$$

Let $\Lambda$ be given by

$$\Lambda = \lim_{t\to\infty} \left(M\tilde{M}\right)^{1/2t}, \tag{5}$$

where $\tilde{M}$ denotes the matrix transpose of $M$. The Lyapunov exponents then equal the logarithm of the eigenvalues of $\Lambda$ [1].

It is easy to show that $M$ satisfies the equation of motion (See, for example, Ref. [10])

$$\frac{dM}{dt} = JSM, \tag{6}$$

where $S$ denotes the symmetric matrix given by

$$H(\mathbf{Z}, t) = \frac{1}{2}\sum_{i,j=1}^{2m} S_{ij} Z_i Z_j, \tag{7}$$

and where

$$J = \begin{pmatrix} 0 & \mathbf{1} \\ -\mathbf{1} & 0 \end{pmatrix}. \tag{8}$$

Here $\mathbf{1}$ denotes the $m \times m$ identity matrix. It follows that the evolution of $M\tilde{M}$ is governed by the equation

$$\frac{d}{dt}M\tilde{M} = JSM\tilde{M} - M\tilde{M}SJ. \tag{9}$$

Standard methods for obtaining the Lyapunov exponents deal with $M$, which is not real symmetric (hence the need for reorthogonalization) and which has exponentially growing elements. To avoid these difficulties we exploit the fact that $M$ is symplectic by making use of the exponential representation of symplectic matrices [6]: Any symplectic matrix $M$ can be written in the form

$$M = \mathrm{e}^{JS_a}\mathrm{e}^{JS_c} \tag{10}$$

where $S_a$ is a symmetric matrix that anticommutes with $J$ and $S_c$ is another symmetric matrix that commutes with $J$. It is important to note that the second matrix on the right hand side of (10) is in fact unitary, so that

$$M\tilde{M} = \mathrm{e}^{2JS_a}. \tag{11}$$

Note that this matrix has fewer degrees of freedom than $M$ and its eigenvectors are orthogonal. Rather than attempting to directly integrate Eqn. (9) which would still have a large numbers problem, we focus our attention on the exponent $JS_a$ in Eqn. (10). It is clear that we now do not have a large numbers problem since $JS_a$ already appears as an exponent.

To proceed further, one obvious approach is to use an explicit representation of $\exp(JS_a)$. Such a representation is well known for $Sp(2)$ and has recently been found for $Sp(4)$ (generalizations to $Sp(2m)$ are in progress) [9]. For the purposes of this Letter we restrict ourselves to one spatial dimension, where the most general symplectic matrix can be written in the form

$$\begin{aligned} M &= \mathrm{e}^{JS_a}\mathrm{e}^{JS_c} \\ &= \mathrm{e}^{\mu(B_2\cos a + B_3 \sin a)}\mathrm{e}^{bB_1}, \end{aligned} \tag{12}$$

where the $B_i$ are basis elements of the Lie algebra $sp(2)$ and where $a$, $b$ and $\mu$ are real coefficients [6]. It follows that

$$M\tilde{M} = \mathrm{e}^{2\mu(B_2\cos a + B_3\sin a)}. \tag{13}$$

Thus, we obtain,

$$\Lambda = \lim_{t\to\infty} \mathrm{e}^{(\mu/t)(B_2\cos a + B_3 \sin a)}. \tag{14}$$

Finally, it is easily shown that the eigenvalues of this matrix are $\mathrm{e}^{\pm\mu/t}$. The Lyapunov exponents are then equal to $\pm\mu/t$ in the limit $t\to\infty$. With this convenient choice of variables, the explicit representation of $M\tilde{M}$ is given by

$$M\tilde{M} = \begin{pmatrix} \cosh 2\mu + \sin a \sinh 2\mu & \cos a \sinh 2\mu \\ \cos a \sinh 2\mu & \cosh 2\mu - \sin a \sinh 2\mu \end{pmatrix}. \tag{15}$$

The unknown quantities $a$ and $\mu$ can grow in time at most as $O(t)$. We can obtain differential equations for these quantities by returning to Eqn. (9), the dynamical equation for $M\tilde{M}$.

For simplicity, we will assume that $H$ contains no term proportional to $qp$, so that the matrix $S$ in (7) is of the form

$$S = \begin{pmatrix} s_{11} & 0 \\ 0 & s_{22} \end{pmatrix}. \tag{16}$$

After some manipulation, Eqns. (7)–(9) lead to the following:

$$\begin{aligned}\frac{d\mu}{dt} &= \frac{1}{2}(s_{22}-s_{11})\cos a,\\ \frac{da}{dt} &= s_{11}+s_{22}-(s_{22}-s_{11})\sin a\coth\mu.\end{aligned} \tag{17}$$

From the initial condition $M(0) = I$, if we choose $\mu(0) = 0$, then $\cos^2 a(0) = 1$, *i.e.*, $a(0) = 0$ or $\pi$. These differential equations form the basis of our method for calculating the Lyapunov exponents of Hamiltonian systems: They are stepped forward in time numerically till some desired convergence for the exponents, $\pm\mu/t$, is achieved. Later we will also show how to apply the method to certain non-Hamiltonian systems.

As a first concrete example, we consider the newly developed "core-halo" model which describes beam halo in mismatched charged particle beams [11]. The transverse equation of motion for a halo particle in this model, assuming constant external focusing, is

$$\ddot{x} + x - (1-\eta^2)f(x, r(t)) = 0 \tag{18}$$

where $x$ is the position variable for a halo particle, $f(x, r(t))$ is the force due to the space charge of the beam core, and $r(t)$ is the time dependent *rms* radius of the core. The core radius is assumed to follow the envelope equation

$$\ddot{r} + r - \frac{1-\eta^2}{r} - \frac{\eta^2}{r^3} = 0. \tag{19}$$

Here units have been chosen so that the time independent solution of (19) (*i.e.*, a matched beam) is given by $r = 1$. In these units $\eta = 0$ corresponds to the space charge dominated regime, while $\eta = 1$ corresponds to the emittance dominated regime. We now assume

$$f(x, r) = \frac{x}{x^2 + r^2} \tag{20}$$

which has the correct asymptotic behavior: the force is linear when $x \ll r$, and it is inversely proportional to $x$ when $x \gg r$. The Eqns. (18) and (19) describe a driven nonlinear system with a mixed phase space as demonstrated by the stroboscopic plot shown in Fig. 1. The presence of a chaotic band is important because particles initially in the core can leak through the broken separatrix and be carried to large amplitudes. The presence of such large amplitude particles can cause unacceptably high radioactivation levels in high intensity linacs planned for future accelerator-driven technologies [12]. Leakage through the separatrix can be enhanced through particle collisions and recent work has shown that this rate is controlled by the Lyapunov exponent [13]. We have computed the Lyapunov exponent for this system by integrating (17) with

$$
\begin{aligned}
s_{11} &= 1-\left(1-\eta^2\right)\left(\frac{1}{x_0^2+r^2}-\frac{2x_0^2}{\left(x_0^2+r^2\right)^2}\right)\\
s_{22} &= 1
\end{aligned}
\tag{21}
$$

where $x_0$ denotes the fiducial trajectory. Fig. 2 displays our result for the Lyapunov exponent against time. The slow convergence of the exponent is typical of Hamiltonian systems.

So far we have dealt with explicitly Hamiltonian systems. However, the only real requirement for using our method is that the linearized deviation equations in *some* variables be Hamiltonian. This allows for the inclusion of damped systems in our scheme. As an example, we now consider the following general driven nonlinear equation

$$\ddot{x}+\lambda\left(1-\epsilon x^2\right)\dot{x}+V'(x)=a\cos(\omega t). \tag{22}$$

By appropriate choices of $\lambda$, $\epsilon$, and $V(x)$, this reduces to an assortment of well-known equations including van der Pol ($\lambda<0$, $\epsilon=1$, $V(x)=(1/2)x^2$), Duffing ($\lambda>0$, $\epsilon=0$, $V(x)=\alpha x^2+\beta x^4$), and the damped driven pendulum ($\lambda>0$, $\epsilon=0$, $V(x)=1-\cos(x)$). In terms of the deviation variable $\delta$, the linearization of (22) yields

$$\ddot{\delta}+\lambda\left(1-\epsilon x_0^2\right)\dot{\delta}+\left(V''(x_0)-2\epsilon\lambda x_0\dot{x}_0\right)\delta=0 \tag{23}$$

where $x_0$ represents the fiducial trajectory. Introducing the new variable $\Delta$ defined through

$$\Delta=\delta e^{-g(t)} \tag{24}$$

where

$$\dot{g}=-\frac{1}{2}\lambda\left(1-\epsilon x_0^2\right), \tag{25}$$

Eqn. (23) reduces to that describing an undamped oscillator with time dependent frequency,

$$\ddot{\Delta}+\left(V''(x_0)-\epsilon\lambda x_0\dot{x}_0-\frac{1}{4}\lambda^2\left(1-\epsilon x_0^2\right)^2\right)\Delta=0. \tag{26}$$

It is now straightforward to proceed using our method: for linear damping ($\epsilon=0$) the Lyapunov exponents $\chi_\pm$ of this system are given by

$$\chi_\pm=-\frac{1}{2}\lambda\pm\lim_{t\to\infty}\frac{1}{t}\mu_0(t) \tag{27}$$

where $\mu_0$ follows from solving (17) for the system defined by (26). When $\epsilon\neq0$,

$$\chi_\pm=\lim_{t\to\infty}\frac{1}{t}\left(g(t)\pm\mu_0(t)\right) \tag{28}$$

modulo terms that are exponentially suppressed at late times. Figs. 3(a) and 3(b) show the Lyapunov exponents of the van der Pol system calculated using our method. For the chosen set of parameters our results are in agreement with those of Ref. [4]. In contrast with the results shown in Fig. 2, the convergence of the exponents is much faster, as is typical of non-weakly damped systems.

In summary we have described a method for computing Lyapunov exponents that exploits the underlying symplectic structure of Hamiltonian dynamics. Just as symplectic integrators are not a panacea for all time integration problems, we do not expect our method to have universal applicability and advantages. However, when applicable, the method has certain advantages over standard techniques, most importantly the lack of systematic errors associated with intermittent reorthogonalization and rescaling.

The authors acknowledge useful discussions with Alex Dragt, Michael Mattis, and Michael Nieto. This work was supported by the U. S. Department of Energy at Los Alamos National Laboratory and by the Air Force Office of Scientific Research.

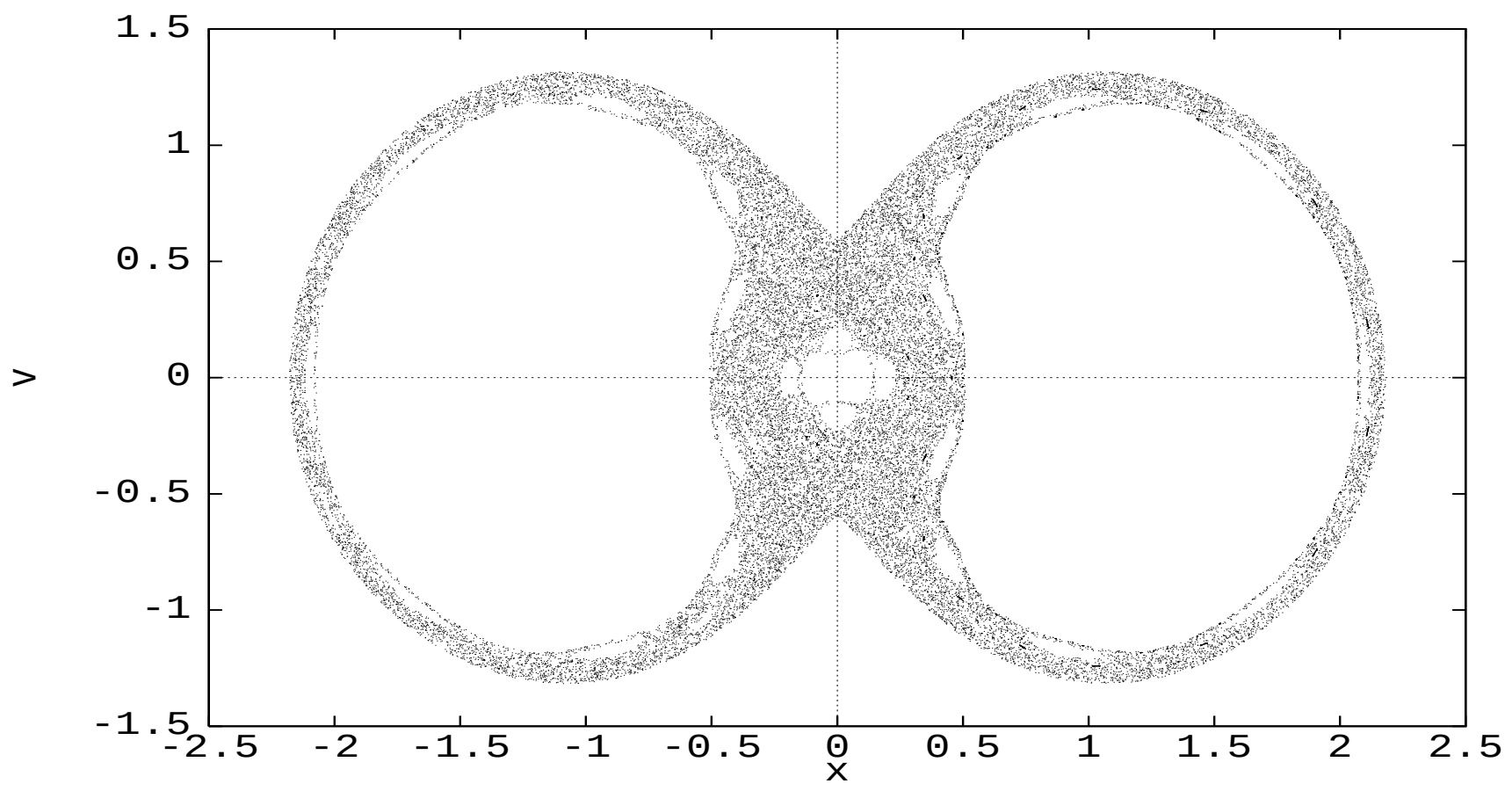

Figure 1: Stroboscopic plot of the chaotic sea in the core-halo model. Snapshots were taken at successive beam minima for 32 test particles. Parameter values were $r(0) = 0.6$, $\dot{r}(0) = 0$, and $\eta = 0.2$.

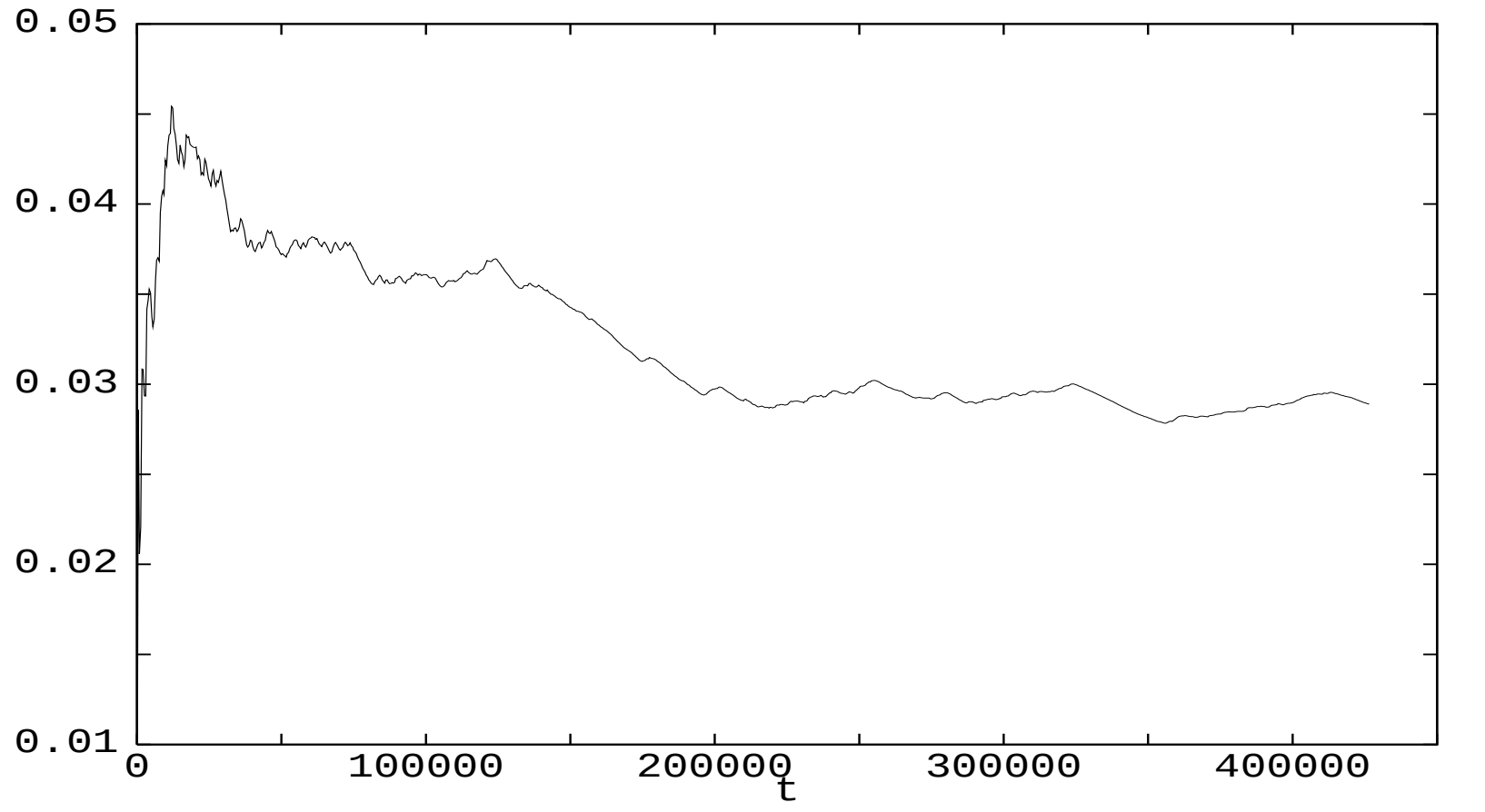

Figure 2: Positive Lyapunov exponent for the core-halo model in a typical run. Parameters are the same as in Fig. 1. The simulation was run for $10^5$ periods of the driving force.

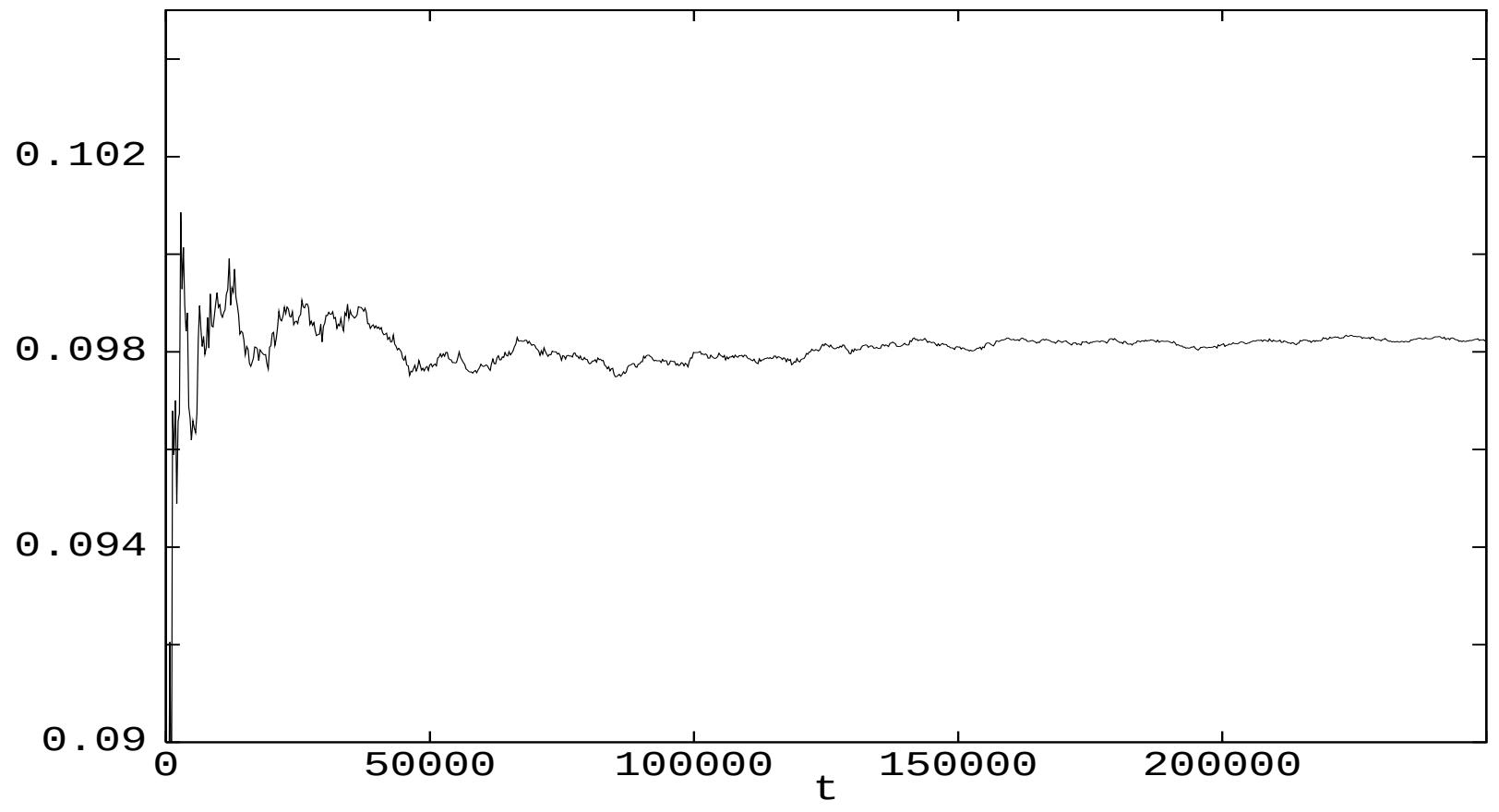

Figure 3: (a) Positive Lyapunov exponent for the van der Pol oscillator with parameters $\lambda = -5$, $a = 5$, and $\omega = 2.466$ (taken from Ref. [4]). The total time for this run corresponds to approximately $10^5$ periods of the driving force.

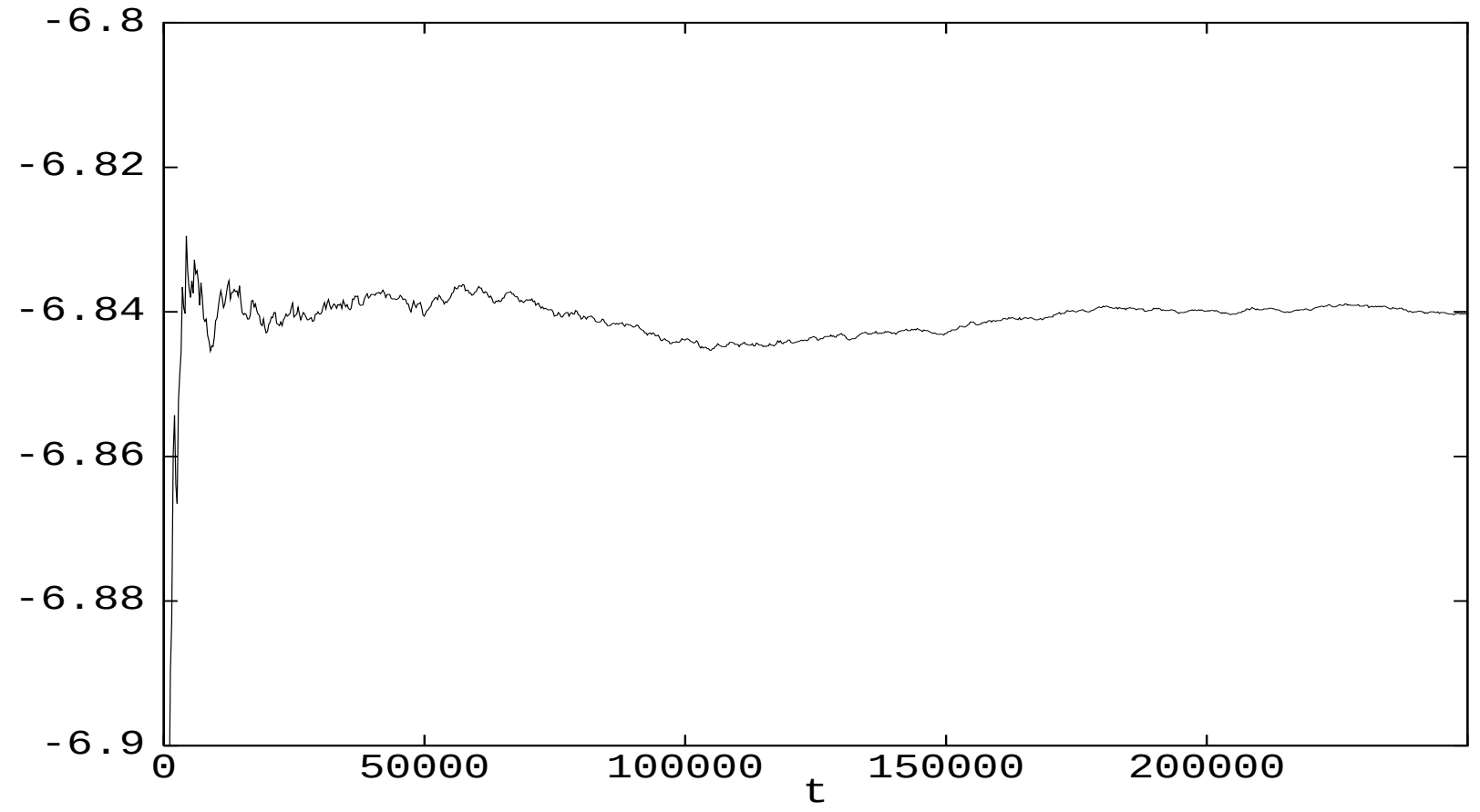

Figure 3: (b) Negative Lyapunov exponent for the van der Pol oscillator with the same parameters as in Fig. 3(a).